\def\frac#1#2{{\textstyle{#1\over #2}}}
\def\hf{\frac12}

\documentstyle[preprint,prb,aps]{revtex}
\begin{document}
\draft
\title{The Mass Gap of the Nonlinear $\sigma$ Model\\ through the Finite
Temperature Effective Action}
\author{David S\'en\'echal}
\address{ Centre de Recherche en Physique du Solide et D\'epartement de
Physique,}
\address{Universit\'e de Sherbrooke, Sherbrooke, Qu\'ebec, Canada J1K 2R1.}
\date{November 1992}
\maketitle
\begin{abstract}
The $O(3)$ nonlinear $\sigma$ model is studied in the disordered phase,
using the techniques of the effective action and finite temperature field
theory. The nonlinear constraint is implemented through a Lagrange
multiplier. The finite temperature effective potential for this multiplier
is calculated at one loop. The existence of a nontrivial minimum for this
potential is the signal of a disordered phase in which the lowest excited
state is a massive triplet. The mass gap is easily
calculated as a function of temperature in dimensions 1, 2 and 3.
In dimension 1, this gap is known as the Haldane gap, and its temperature
dependence is compared with experimental results.
\end{abstract}

The $O(3)$ nonlinear $\sigma$ model (NL$\sigma$) offers a good description of
quantum antiferromagnets at long wavelengths, as suggested by renormalization
group analyses \cite{CHN}. The mapping between the nonlinear $\sigma$ model and
the Heisenberg model was first established by Haldane \cite{HaldaneA},
who also argued that a topological Hopf term should be present in the case
of half-integer spins in one dimension \cite{HaldaneB} (this distinction
between half-integer and integer spins disappears in two dimensions).
For integer spins, in which case the Hopf term is absent, Haldane argued that
the lowest excitations should exhibit a mass gap. Accordingly, these
excitations (the magnons) are no longer Goldstone modes and the rotation
symmetry of the spins is no longer spontaneously broken: the symmetry has
been dynamically restored by quantum fluctuations. This phenomenon has
found an experimental realization in quasi one-dimensional antiferromagnets
in their disordered phase \cite{BuyersA}.

In this paper we calculate the temperature dependence of this mass gap in
dimensions 1 to 3 within the framework of the finite temperature effective
action of the nonlinear $\sigma$ model. The use of this technique is
facilitated by the Lorentz invariance of the NL$\sigma$ model, which ensures
that the magnon speed is not affected by quantum corrections.

The $O(3)$ nonlinear $\sigma$ model describes the dynamics of a unit vector
${\vec\varphi}$ representing the local direction of the staggered
magnetization. The Euclidian action obtained from the Heisenberg model is
\begin{equation}
S~=~ {v\over 2g}\int d\tau d^dx~\left\{ {1\over v^2}
(\partial_\tau{\vec\varphi})^2+ (\partial_i{\vec\varphi})^2\right\}
\end{equation}
wherein $v$ is the spin-wave velocity and $g$ a coupling constant.
the constraint ${\vec\varphi}^2=1$ is understood.
If we choose a set of units in which the speed $v$ is
unity, implement the constraint ${\vec\varphi}^2=1$ by a Lagrange multiplier
$\sigma$, and rescale the field ${\vec\varphi}$ by $\sqrt{g}$ in order to bring
its kinetic term into the standard form, we obtain the equivalent action
\begin{equation}
S~=~ \int d\tau d^dx~\big\{\frac12 \partial_\mu{\vec\varphi}
\partial_\mu{\vec\varphi} - \frac12\sigma({\vec\varphi}^2 - 1/g) \big\}
\label{actionA}
\end{equation}
wherein the index $\mu$ goes from 0 to $d$. The spin-wave velocity $v$
may always be restored later using dimensional analysis.

The action (\ref{actionA}) is symmetric with respect to global rotations of
the field ${\vec\varphi}$, which is the relevant order parameter. The
degenerate classical ground states are configurations in which the field
${\vec\varphi}$ is uniform. Such ground states break the full symmetry group
$O(3)$ down to $O(2)$. Goldstone's theorem would then imply the existence of
two massless modes, corresponding to the two magnon polarizations. However, the
quantum ground state, to which Goldstone's theorem refers, does not break the
$O(3)$ symmetry if we are in a disordered phase. The symmetry of the
quantum ground state may be determined by calculating the one-loop
effective potential $V(\sigma)$ for the field $\sigma$, following the
discussion by Coleman\cite{Coleman}. If this potential has a global minimum at
some nonzero value $\sigma_0$ of the field $\sigma$,  then we must expand the
field $\sigma$ around its ground state value $\sigma_0$. The latter then gives
rise to a mass term $\hf\sigma_0{\vec\varphi}^2$ in the action (\ref{actionA}).
As a consequence, the lowest lying excitations are no longer a pair
of massless Goldstone modes (the infrared instability has been removed)
but a massive triplet: the elementary
excitations of the three-component field ${\vec\varphi}$ with mass gap
$\Delta=\sqrt{\sigma_0}$.

The one-loop effective potential $V(\sigma)$
is obtained by summing an infinite sequence of Feynman graphs, each with
$n$ insertions of $\sigma$ and $n$ internal ${\vec\varphi}$ lines.
This amounts to evaluating the partition function with $\sigma$ constant, which
is done simply by Gaussian integration. The result of this
standard calculation is
\begin{equation}
V(\sigma) = {\sigma\over 2g} -3\int{dp_0\over (2\pi)}
\int{d{\vec p}\over (2\pi)^d}~\log\left\{ 1+ {\sigma\over p_0^2+
{\vec p}^2}\right\} \end{equation}
where we have included the tree-level potential $\sigma/2g$ (we use the
convention $\hbar=1$ throughout).
In order to find out whether $V(\sigma)$ has a minimum, we consider instead
the derivative
\begin{equation}
V'(\sigma) = {1\over 2g} -3\int{dp_0\over (2\pi)}\int{d{\vec p}\over
(2\pi)^d} ~{1\over p_0^2 +{\vec p}^2+ \sigma}
\end{equation}
The extension of this formula to a finite temperature $T$ is obtained by
replacing the integration over $p_0$ by a sum over Matsubara
frequencies $\omega_n = 2\pi nT$:
\begin{equation}
\int {dp_0\over 2\pi}f(p_0) ~\rightarrow~ T\sum_n f(\omega_n)
\end{equation}
Using the standard result ($\beta=1/T$)
\begin{equation}
\sum_n {1\over \omega_n^2 + z^2} ~=~ {\beta\over 2z}{\rm coth}(\beta z/2)
\end{equation}
we may write
\begin{equation}
2V'(\sigma) = {1\over g} -3\int {d{\vec p}\over (2\pi)^d}~
{1\over\sqrt{{\vec p}\kern0.8pt^2 + \sigma}}
{\rm coth}\left(\hf\beta\sqrt{{\vec p}\kern0.8pt^2 + \sigma}\right)
\label{derivA}
\end{equation}
At zero temperature, the above reduces to
\begin{equation}
2V'(\sigma) = {1\over g} -3K_d\int_0^\Lambda ~
{p^{d-1}dp\over\sqrt{p^2 + \sigma}}
\label{derivB}
\end{equation}
where $K_1^{-1}=2\pi$ and $K_d^{-1} = 2^{d-1}\pi^{d/2}\Gamma(d/2)$ if $d>1$.
An ultraviolet cutoff $\Lambda$ has been introduced. We point out that the
NL$\sigma$ model is a renormalizable field theory only in $d=1$; this means
that our treatment for $d>1$ will only be valid in the limit of large
distances. Without taking $\Lambda$ to infinity, we will consistently neglect
terms of order $\Lambda^{-1}$ or more. This being understood, finite
temperatures will not introduce any new dependence on $\Lambda$: Indeed, the
difference  $V_\delta \equiv V(\sigma,T)-V(\sigma,0)$ is given by
\begin{equation}
\label{derivC}
2V_\delta'(\sigma) = -3K_d\int_0^\Lambda {p^{d-1}dp\over\sqrt{p^2
+\sigma}}\left\{{\rm coth}\left(\hf\beta\sqrt{p^2 + \sigma}\right) - 1\right\}
\end{equation}
Since the function ${\rm coth} x-1$ decreases exponentially\cite{noteA}
as $x\to\infty$, the
above integral is UV finite and the cut-off may be removed.
The dependence on $\Lambda$ is contained solely in $V(\sigma,0)$.
A change of integration variables yields
\begin{equation}
2V_\delta'(\sigma) = -T^{d-1} f_d(y)
\label{derivD}
\end{equation}
where $y=\sqrt{\sigma}/T$ and
\begin{equation}
f_d(y) = 3y^{d-1}K_d\int_1^\infty dz~{{\rm coth}(\hf yz) - 1\over
(z^2-1)^{1-d/2}}
\end{equation}
We find that
\begin{mathletters}
\label{fform}
\begin{eqnarray}
f_1(y) &=& {3\over\pi}\sum_{n=1}^\infty K_0(ny)\\
f_2(y) &=& {3\over\pi}\left\{\frac12 y - \log(2{\rm sinh}(y/2))\right\}\\
f_3(y) &=& {3y\over\pi^2}\sum_{n=1}^\infty {1\over n}K_1(ny)
\end{eqnarray}
\end{mathletters}
($K_0$ and $K_1$ are modified Bessel functions.)

Let us now look in more detail at specific cases, starting with $d=1$. This
has been done explicitly by Coleman\cite{Coleman} for the $CP^1$ model,
which is equivalent to the $O(3)$ nonlinear $\sigma$ model. From Eq.
(\ref{derivB}) we obtain, at zero temperature,\cite{noteB}
\begin{equation}
2V'(\sigma) = {1\over g}+{3\over 2\pi}\log\left(\sigma/\Lambda^2\right)
\qquad (d=1)
\end{equation}
We see that the equation $V'(\sigma)=0$ always has a solution, at the point
$\sigma_0 = \Lambda^2\exp-(2\pi/3g)$, which implies a zero temperature Haldane
gap $\Delta_0=\Lambda\exp-\pi/3g$.

{}From Eqs. (\ref{derivD}) and (\ref{fform}), the finite temperature
result is then \begin{equation}
2V'(\sigma) = {1\over g}+{3\over 2\pi}\log\left(\sigma/\Lambda^2\right) -
{3\over\pi}\sum_{n=1}^\infty K_0(n\sqrt{\sigma}/T)
\end{equation}
The finite temperature Haldane gap $\Delta=\sqrt{\sigma_0}$ is therefore
obtained through the solution of the equation
\begin{equation}
\hf\log(\delta) = \sum_{n=1}^\infty K_0(n\delta/t)
\label{oneD}
\end{equation}
where we have defined the reduced gap $\delta=\Delta/\Delta_0$ and the
reduced temperature $t=T/\Delta_0$. This gap equation may be solved
numerically. The solution is shown on Fig.1. The gap is seen to maintain its
$T=0$ value until about $T\sim 0.2\Delta_0$, after which it takes off linearly.
We naturally expect the gap to grow with temperature since $\Delta$, being the
inverse of the correlation length, is in some sense a measure of disorder. This
temperature dependence seems confirmed by experiments\cite{BuyersB}.

The dependence of $\Delta$ on $T$ is not qualitatively
affected by the introduction of anisotropy in the $\sigma$ model. Easy-axis
anisotropy is introduced via a mass term $\hf M^2\varphi_z^2$
in the action (\ref{actionA}), while easy-plane anisotropy
is rendered by a mass term $\hf M^2(\varphi_x^2+\varphi_y^2)$.
The above calculation of the Haldane gap is easily adapted to the
anisotropic case by shifting $\sigma$ by $\alpha$ for one (two) out of three
components of ${\vec\varphi}$ in the easy-axis (easy-plane) case. The gap
equation (\ref{oneD}) is thus modified into
\begin{eqnarray}
\frac14\log\left\{\left({\mu^2+\delta^2\over\mu^2+1}\right)\delta^4\right\}
=\sum_{n=1}^\infty K_0(n\sqrt{\delta^2+\mu^2}/t)+
2\sum_{n=1}^\infty K_0(n\delta/t)
\end{eqnarray}
where we have defined the reduced anisotropy $\mu=M/\Delta_0$ and considered
the easy-axis case. Even if the value of $\Delta_0$ is affected by the
anisotropy, everything else being fixed, we have checked that the $t$
dependence of the reduced gap $\delta$ is not changed appreciably by the
introduction of an anisotropy even as large as $\mu=1$.
However, an easy-axis anisotropy splits the Haldane triplet into a doublet with
gap $\Delta_1=\Delta$ and a singlet with gap $\Delta_2=\sqrt{\Delta^2+M^2}$.
The weighted average of the two gaps, normalized to its zero temperature value,
is shown on the lower curve of Fig.1 for $\mu=1.9$. This value of $\mu$,
appropriate for NENP, is obtained from the observed values of $\Delta_1$ (14K)
and $\Delta_2$ (30K)\cite{Renard}.

We next consider the two-dimensional case. At $T=0$ this was studied by
Deser and Redlich\cite{Deser} for the $CP^1$ model. The result of the
integration (\ref{derivB}) is
\begin{equation}
2V'(\sigma) = {1\over g}-{3\Lambda\over 2\pi} +
{3\sqrt\sigma\over 2\pi}\qquad (d=2)
\end{equation}
The interesting feature here is the appearance of a critical coupling
$g_c= 2\pi/\Lambda$ below which the gap equation $V'(\sigma)=0$ has no
solution. This is the signal of a phase transition in coupling space: below
$g_c$, the $O(3)$ symmetry is spontaneously broken and massless Goldstone modes
appear. At finite temperatures we have, following Eq. (\ref{fform}):
\begin{equation}
2V'(\sigma) = {1\over g}-{1\over g_c} +
{3T\over\pi}\log\left[2{\rm sinh}(\sqrt\sigma/2T)\right]
\end{equation}
The gap equation has now a solution for all nonzero values of $T$,
indicating that the ordered phase does not survive at finite temperatures,
as is already known from RG calculations\cite{CHN}. Let us define the
difference $\eta=1/g_c-1/g$. The solution of the gap equation is then
\begin{equation}
\Delta~=~ 2T{\rm sinh}^{-1}\left[\hf\exp(\pi\eta/3T)\right]
\label{gapTwo}
\end{equation}
If $\eta>0$ (disordered phase) the gap tends towards the constant
$2\pi\eta/3$ as $T\to0$, i.e., $\eta=3\Delta_0/2\pi$.
If, on the contrary, $\eta<0$ (ordered phase) the gap behaves like
$\Delta\approx T\exp-\pi|\eta|/3T$ as $T\to0$, i.e., it vanishes
faster than any power law.
At the critical value $\eta=0$ the gap vanishes linearly as $T$ is
lowered:
\begin{equation}
\Delta= 2T{\rm sinh}^{-1}(\hf) \approx 0.962~T
\end{equation}
These results are illustrated in Fig.2. Similar results were obtained within
the RG analysis of Ref.[\onlinecite{CHN}] in terms of the correlation length
$\xi=1/\Delta$. Experimental results are available in the ordered phase
($\eta<0$) and compare well with the theory (see Ref.[\onlinecite{CHN}]).
To the author's knowledge, the disordered phase ($\eta>0$) has not been
observed yet. Although it is accepted that this phase does not exist at $T=0$
for the Heisenberg model with spin $s\geq 1$, nothing certain can be said
about the spin-$\hf$ case.

Let us point out that the cross-over region, i.e., the temperature range
over which thermal fluctuations overtake quantum fluctuations, is
characterized by the condition $\Delta\sim T$ which, on the phase diagram
$(g,T)$, is met on the curve $T\sim25.3\eta$, according
to Eq. (\ref{gapTwo}). When $T$ goes beyond
this value the states over the gap are thermally excited or, stated
differently, the correlation length $\xi$ exceeds the extent $\beta$ of the
Euclidian time direction; the fluctuations then happen mostly in the
spatial directions.

Let us finally discuss the three-dimensional case, first at $T=0$. The
integration (\ref{derivB}) yields
\begin{equation}
2V'(\sigma) = {1\over g}-{3\Lambda^2\over 4\pi^2} + {3\sigma\over 8\pi^2}
\left\{ \log(2\Lambda^2/\sigma)-1\right\}
\label{threeD}
\end{equation}
Again, we have dropped terms of order $1/\Lambda$.
It is understood that $\sigma$ is much smaller than the cutoff $\Lambda$, so
that $V'(\sigma)$ is increasing monotonically with $\sigma$. Like in dimension
2, there is a critical coupling $g_c=4\pi^2/3\Lambda^2$ below which the
symmetry is spontaneously broken at $T=0$. The difference with $d=2$ is that
this ordered phase survives at finite temperature. Indeed, following
Eqs. (\ref{fform}) and (\ref{threeD}), the gap equation is
\begin{equation}
0 = {1\over g}-{1\over g_c} + {3\Delta^2\over 8\pi^2}
\left\{ \log(2\Lambda^2/\Delta^2)-1\right\} - T^2f_3(\Delta/T)
\end{equation}
The function $f_3(y)$ is monotonically decreasing, rather like an
exponential, starting with $f_3(0)\approx0.5$.
If the above equation has no solution at $T=0$ ($g<g_c$), then an
increase in temperature will bring a solution at some point, since the
r.h.s. can only decrease as $T$ increases.
The transition to the ordered phase occurs when $\Delta=0$;
the relation between $g$ and $T$ that follows is
\begin{equation}
{1\over g}-{1\over g_c} = T^2f_3(0)
\end{equation}
We see that no matter how small $g$ may be below $g_c$, a critical
temperature  exists beyond which we recover a disordered phase.
However, this result is not to be trusted at arbitrary low $g$, since
we tacitly assumed that $T\ll\Lambda$.

If we place ourselves above $g_c$, denote by $\Delta_0$ the zero temperature
gap and define reduced variables $\delta=\Delta/\Delta_0$, $t=T/\Delta_0$ and
$A=\log(\Lambda/\Delta_0)+\hf(\log2-1)$, the gap equation becomes
\begin{equation}
A(1-\delta^2)-\delta^2\log\delta + {4\pi\over3}t^2 f_3(\delta/t)=0
\end{equation}
Its solution for $\delta$ as a function of $t$ may be found numerically,
and is shown on Fig. 3 for two values of $A$. Note that we cannot rid
ourselves of this extra parameter, but its logarithmic dependence on the
cut-off causes only a mild non universality.
At the critical value $g=g_c$, we find that the gap vanishes linearly with
$T$, like in dimension 2, albeit with mild corrections.

In conclusion, calculating the one-loop effective potential for the
Lagrange multiplier $\sigma$ yields the Haldane gap as a function of
temperature in a rather direct way. The solution is compatible with
experimental results. The extension to two and three dimensions is
straightforward although, to the author's knowledge, experimental
results in the quantum disordered regime are then lacking. Since
no rigorous argument has yet been found that prevents the existence of a
quantum disordered phase in the two-dimensional spin-$\hf$ Heisenberg model,
the question remains open.

\acknowledgements
Stimulating conversations with C. Bourbonnais, W.J.L. Buyers, M. Plumer,
Y. Trudeau and especially A.-M.S. Tremblay are gratefully acknowledged. This
work is supported in part by le Fonds pour la Formation de Chercheurs
et l'Aide \`a la Recherche du Gouvernement du Qu\'ebec (F.C.A.R.).%
%

\begin{figure}
\caption{Magnitude of the reduced Haldane gap $\delta=\Delta/\Delta_0$ as a
function of $t=T/\Delta_0$ in $d=1$ (upper curve). Experimental results from
Ref.[6] are also shown for CsNiCl$_3$ and NENP (error bars
are not included). The lower curve is the weighted average of the two
gaps in the presence of a $\mu=1.9$ easy-axis anisotropy, appropriate for
NENP.}
\label{figA}
\end{figure}

\begin{figure}
\caption{Magnitude of the mass gap $\Delta$ as a function of temperature
for $d=2$ and three values (indicated) of $\eta$.}
\label{figB}
\end{figure}

\begin{figure}
\caption{Magnitude of the reduced mass gap $\delta=\Delta/\Delta_0$ as a
function of $t=T/\Delta_0$ in $d=3$ for $A=4$ (higher curve) and $A=5$ (lower
curve).} \label{figC}
\end{figure}

\begin{references}
\bibitem[*]{noteA}
The expansion ${\rm coth} x-1=2\sum_{n=1}^\infty \exp-2nx$ has proven useful.
%
\bibitem[**]{noteB} The logarithmic cut-off dependence could be removed by
defining a renormalized coupling : $1/g_r\equiv 2V'(\mu^2)$, where $\mu$ is
a subtraction point. The effective potential would then be well-behaved as
$\Lambda\to\infty$ and $g\to 0$, while $g_r$ stays fixed.
However, since we do not intend to send the cutoff to $\infty$, this
procedure would be of little use to us.
%
\bibitem{CHN} S. Chakravarty, B. I. Halperin and D. R. Nelson,
Phys. Rev. {\bf B39}, 2344 (1989).
%
\bibitem{HaldaneA} F. D. M. Haldane, Phys. Lett. {\bf A93},464 (1983).
%
\bibitem{HaldaneB} F. D. M. Haldane, Phys. Rev. Lett. {\bf 50},1153 (1983).
%
\bibitem{BuyersA} W. J. L. Buyers et al., Phys. Rev. Lett. {\bf 56}, 371
(1986).
%
\bibitem{Coleman} S. Coleman, {\em Aspects of Symmetry: selected Erice
Lectures}, Cambridge University Press (1985), chapter 8.
%
\bibitem{BuyersB} W. J. L. Buyers, Z. Tun, A. Harrison, J. A. Rayne and
R. M. Nicklow, Physica {\bf B180-181}, 222 (1992).
%
\bibitem{Renard} J.-P. Renard et al., J. Appl. Phys. {\bf 63}, 3538 (1988).
%
\bibitem{Deser} S. Deser and A. N. Redlich, Phys. Rev. Lett. {\bf 61}, 1541
(1988).
%
\end{references}
\end{document}